\documentclass[9pt,twocolumn,twoside]{osaLeoTeX}   

\journal{ol} 

\setboolean{shortarticle}{true}

\title{Ptychographic reconstruction of pure quantum states}

\author[1]{M. F. Fernandes}
\author[2]{M. A. Sol\'is-Prosser}
\author[1,*]{L. Neves}

\affil[1]{Departamento de F\'isica, Universidade Federal de Minas Gerais, Belo Horizonte, MG 31270-901, Brazil}
\affil[2]{Departamento de Ciencias F\'isicas, Universidad de La Frontera, Temuco, Chile}
\affil[*]{Corresponding author: lneves@fisica.ufmg.br}

\begin{abstract}
The quantum analogue of ptychography, a powerful coherent diffractive imaging technique, is a simple method for reconstructing $d$-dimensional pure states. It relies on measuring partially overlapping parts of the input state in a single orthonormal basis and feeding the outcomes to an iterative phase-retrieval algorithm for postprocessing. We provide a proof of concept demonstration of this method by determining pure states given by superpositions of $d$ transverse spatial modes of an optical field. A set of $n$ rank-$r$ projectors, diagonal in the spatial mode basis, is used to generate $n$ partially overlapping parts of the input and each part is projectively measured in the Fourier transformed basis. For $d$ up to 32, we successfully reconstructed hundreds of random states using $n=5$ and $n=d$ rank-$\lceil d/2\rceil$ projectors. The extension of quantum ptychography for  other types of photonic spatial modes is outlined. 
\end{abstract}

\setboolean{displaycopyright}{false}

\begin{document}

\maketitle

The quantum state fully determines the measurable properties of a physical system. Its knowledge is, thereby, crucial for applications ranging from quantum metrology to quantum information and computation \cite{NielsenBook}. In this sense, techniques of state reconstruction have become fundamental tools in quantum physics. A standard technique, called quantum tomography, relies on projections in a complete set of incompatible bases and a postprocessing algorithm to estimate the state \cite{James01}. In the face of increasing complexity of tomography with the state-space dimension \cite{Haffner05}, recent effort has been directed to simplify the process using prior information \cite{Gross10,Heinosaari13}. In particular, for a quantum state known to be pure, a number of tomographic techniques has been developed showing that up to five measurement bases will determine it in any finite dimension \cite{Goyeneche08,Goyeneche15,Carmeli16,Stefano17,Stefano19}. Although, in practice, purity is just an approximation, in many scenarios it is good enough. Therefore, a robust tomographic scheme for pure states must be resilient to noise in order to deal with this approximate purity.

Recently, two of us have proposed a quantum analogue of \emph{ptychography} as a simple and robust method to reconstruct pure states \cite{Fernandes19}. Ptychography \cite{Faulkner04} is a form of coherent diffractive imaging where a localized illumination is shifted through partially overlapping parts of an object and generates multiple diffraction patterns; these patterns are processed by an iterative phase retrieval algorithm (e.g., the ptychograhic iterative engine (PIE) \cite{Faulkner04,Rodenburg04}), which recover the complex-valued object transmission function. The technique is specially powerful for optical \cite{Thibault08} and electron \cite{Humphry12} microscopy and also found applications in optical encryption \cite{Shi13} and nonlinear imaging \cite{Odstrcil16}. 

In our proposal \cite{Fernandes19}, a $d$-dimensional ($d>2$) pure quantum state is the object of interest and the role of the shifting illumination is played by a set of $n$ rank-$r$ projectors. These projectors, one at a time, ``slice'' the state into $n$ partially overlapping parts and each part is projectively measured in the same orthonormal basis. The generated $nd$ outcomes are then processed by a PIE-based algorithm which gives us an estimate of the state. The method succeeds if the measured data have sufficient diversity and redundancy, which depends on the choice of the projectors (their form, $n$, $r$) and the measurement basis. Restricting to single quantum systems, we have shown that sets with $n$ up to $d$ diagonal projectors of ranks around $d/2$, together with measurements in the Fourier transformed basis, provide successful reconstructions even under noisy conditions.

Comparing with other tomographic approaches for pure states, which require projective measurements in a fixed number of bases \cite{Goyeneche08,Stefano17,Stefano19,Goyeneche15,Carmeli16}, the ptychographic technique \cite{Fernandes19} resorts to a flexible number ($n$) of projective measurements in a single basis. It also uses simpler settings in the experimental setup, since the rank-$r$ projectors are usually easy to implement. Additionally, the reconstruction is done with a simple and fast phase retrieval algorithm. 

In this Letter, we provide a proof of concept demonstration of quantum ptychography by determining pure states given by superpositions of $d$ transverse spatial modes of an optical field. We use rank-$\lceil d/2\rceil$ projectors, diagonal in the spatial mode basis, to generate $n$ partially overlapping parts of the input and each part is projectively measured in the Fourier transformed basis. 
For $d$ up to 32, we reconstructed hundreds of random states using $n=5$ and $n=d$ projectors, achieving high fidelities with fast postprocessing in all cases. Finally, we discuss the straightforward extension of the ptychographic technique for reconstructing other types of spatially encoded photonic states.

\begin{figure}[t!]
\centerline{\includegraphics[width=1\columnwidth]{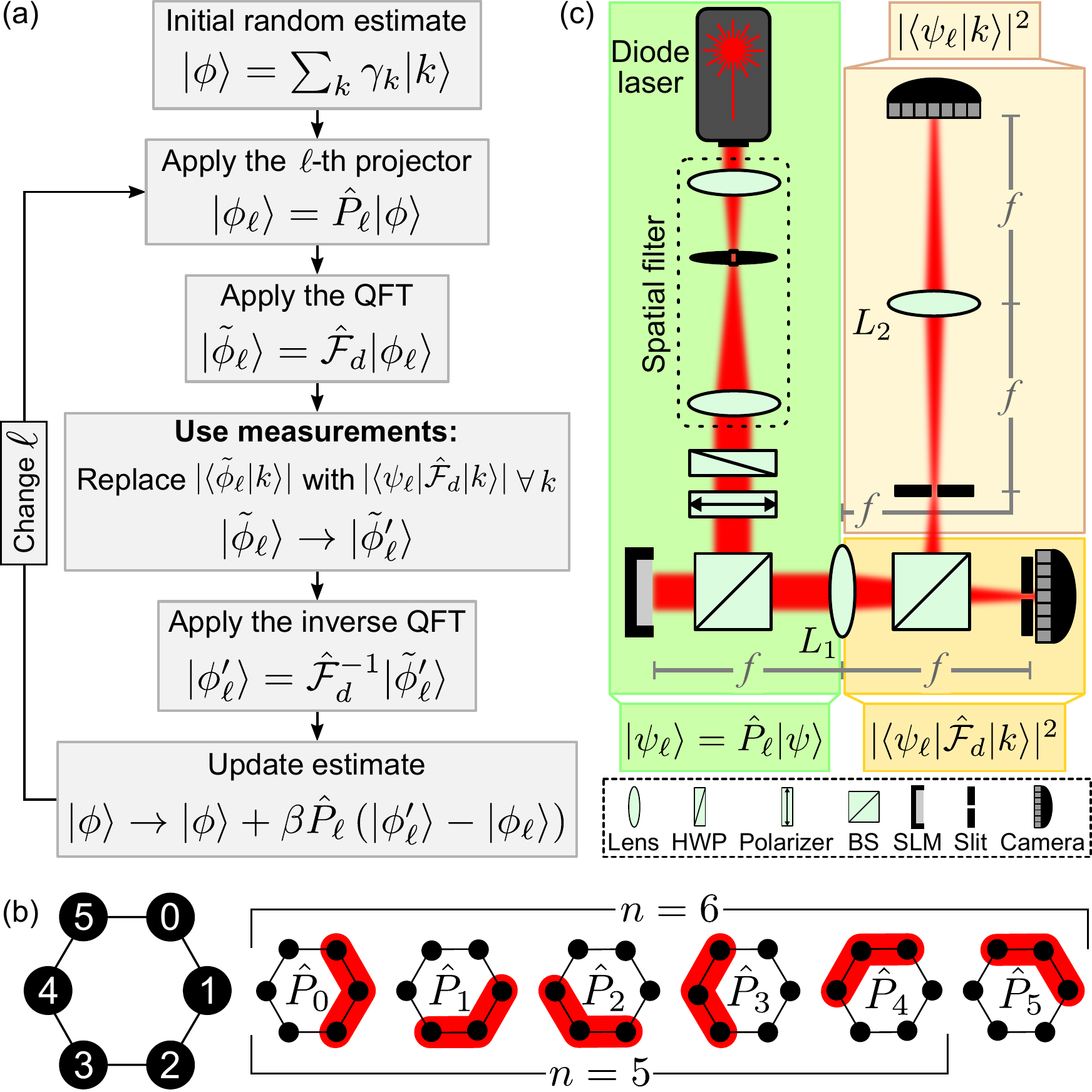}}
\caption{\label{fig:setup} (a) PIE algorithm: to estimate the quantum state from the ptychographic data and the $n$ projectors given by Eq.~(\ref{eq:Proj}), the algorithm is initiated with a random  $|\phi\rangle$ that goes through the steps shown in the diagram. This state is updated and the process is iterated through the closed loop. A single PIE iteration comprises $n$ iterations through this loop, where each $\hat{P}_\ell$ and the corresponding data is used once to update the estimate. At each iteration, the relative distance $D=(\| |\phi_{\rm updated}\rangle-|\phi_{\rm current}\rangle\|/\||\phi_{\rm current}\rangle\|)^2$ between the estimates is computed. We used $\beta=1.6$ to control the step-size of the update (see \cite{Fernandes19}) and made the algorithm terminate after reaching either $D<10^{-2}$ or 25 PIE iterations; in the latter case, we made it restart with a new random estimate and up to 100 restartings were allowed. At the end, a pure state is delivered which must be normalized. (b) Schematic representation of the projectors of Eq.~(\ref{eq:Proj}) for $d=6$ and $r=3$. The red shade highlights the selected levels in $\mathcal{H}_d$. Bottom: family (i); top: family (ii). (c) Experimental setup (see text). The insets show the role of each shaded region in the experiment. Lenses $L_1$ and $L_2$ have focal length $f=30\!$~cm. HWP: half-wave plate; BS: beam splitter; SLM: spatial light modulator.}
\end{figure}

An arbitrary pure quantum state in a $d$-dimensional Hilbert space $\mathcal{H}_d$ can be written as
$|\psi\rangle=\sum_{k=0}^{d-1}c_k|k\rangle,$
where $\{|k\rangle\}_{k=0}^{d-1}$ is an orthonormal basis for $\mathcal{H}_d$ (here called computational basis) and $\{c_k\}_{k=0}^{d-1}$ (with $\sum_k|c_k|^2=1$) the set of complex coefficients which completely specify $|\psi\rangle$. The ptychographic method to determine these coefficients requires a set of $n$ projectors $\{\hat{P}_\ell\}_{\ell=0}^{n-1}$ of rank $r$ ($>1$), where each $\hat{P}_\ell$ has a partial overlap with at least one other partner and all levels of $\mathcal{H}_d$ are addressed at least once \cite{Fernandes19}. After specifying them, the protocol goes as follows: first, one applies the $\ell$-th projector on the input ensemble of quantum systems described by an unknown $|\psi\rangle$. Next, the output sub-ensemble, described by the (unnormalized) state $|\psi_\ell\rangle=\hat{P}_\ell|\psi\rangle$, is measured in the Fourier basis $\{\hat{\mathcal{F}}_d|k\rangle\}_{k=0}^{d-1}$, where $\hat{\mathcal{F}}_d$ is the quantum Fourier transform (QFT) acting on $\mathcal{H}_d$. These steps are repeated for each $\hat{P}_\ell$ and in the end provide a set of $n$ count distributions, proportional to $\{|\langle\psi_\ell|\hat{\mathcal{F}}_d|k\rangle|^2\}_{k=0}^{d-1}$, which are fed into the PIE algorithm that estimates the state. The PIE diagram and its operation is presented in Fig.~\ref{fig:setup}(a).

As shown in \cite{Fernandes19}, a suitable choice for $\{\hat{P}_\ell\}_{\ell=0}^{n-1}$ comprises projectors diagonal in the computational basis given by 
\begin{equation}   \label{eq:Proj}
\hat{P}_\ell=\sum_{j=0}^{r-1}|j\oplus s_\ell\rangle\langle j\oplus s_\ell|,
\end{equation}
where $\oplus$ denotes addition modulo $d$ and $s_\ell$ is a nonnegative integer setting the skip between adjacent $\hat{P}_\ell$'s. Here, we shall use two families of Eq.~(\ref{eq:Proj}) with the following specifications: (i) $n=5$ and $s_\ell=\ell\lfloor d/5\rfloor$; (ii) $n=d$ and $s_\ell=\ell$. For both, $r=\lceil d/2\rceil$. Figure~\ref{fig:setup}(b) illustrates their action for $d=6$. The overlapping condition and the access to all $\mathcal{H}_d$ levels are clearly accomplished. The family (i) uses $5d$ measurement outcomes, which is comparable to other reconstruction techniques \cite{Goyeneche08,Stefano17,Stefano19,Goyeneche15,Carmeli16}; the family (ii) requires $d^2$ outcomes, which is an overcomplete dataset to reconstruct pure states.  

To demonstrate the ptychographic method, we resort to pure states represented by superpositions of $d$ distinct transverse spatial modes of an optical field. These spatial modes are defined by $d$ non-overlapping paths that photons, traveling along $z$, may take in one transverse dimension $x$ (e.g., by using a $d$-slit array \cite{Neves05}). Taking these paths as our computational basis, the rank-$r$ projectors of Eq.~(\ref{eq:Proj}) are implemented simply by blocking $d-r$ of them and leaving the others open. In turn, the projective measurement in the Fourier basis is accomplished by measuring the light intensities at $d$ transverse positions in the plane of the optical Fourier transform of the spatial modes performed by a spherical lens. These positions are \cite{Prosser17} 
\begin{equation}    \label{eq:positions}
x_j=-\lambda f\mu_j/\delta d, 
\end{equation} 
for $j=0,\ldots,d-1$, where $\lambda$ is the light wavelength, $f$ the lens focal length, $\delta$ the separation between adjacent paths, and $\mu_j=j$ if $j\leq d/2$ or $\mu_j=j-d$ if $j>d/2$. Each ``pixel'' detector at $(x_j,f)$ postselects the state $\hat{\mathcal{F}}_d|j\rangle$ \cite{Prosser17}.

\begin{figure*}[t!]
\centerline{\includegraphics[width=1\textwidth]{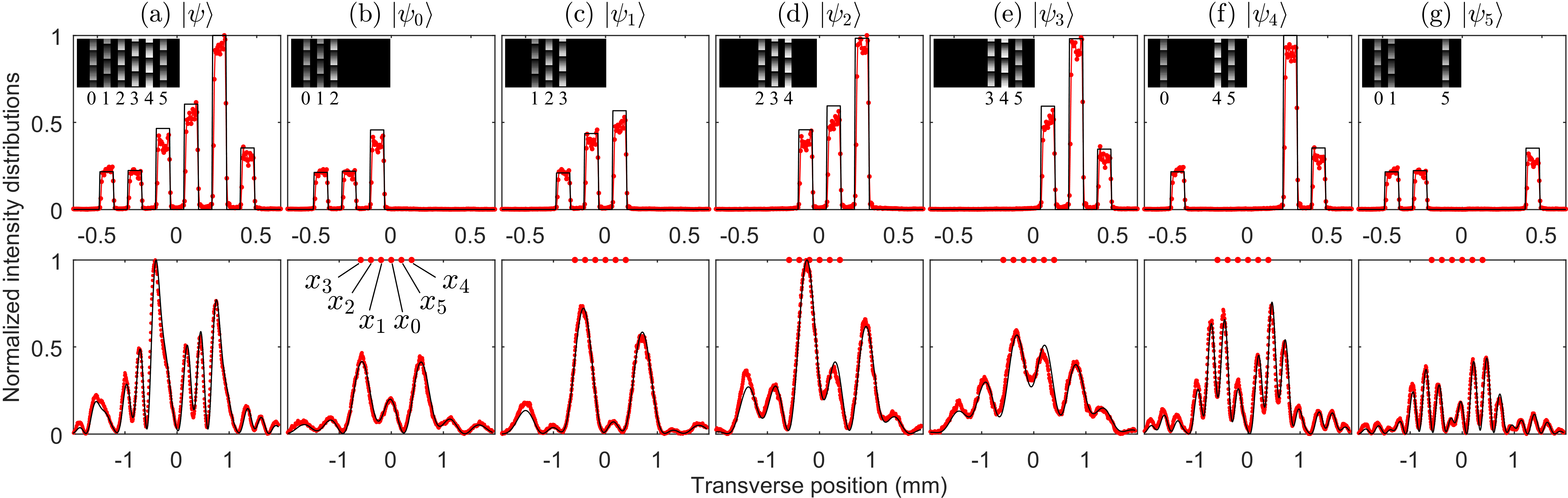}}
\caption{\label{fig:NearFar}  Normalized near- (first row) and far-field (second row) intensity distributions for a 6-dimensional target state $|\psi\rangle$ (a) and the corresponding projected states $|\psi_\ell\rangle$ (b)--(g). Experiment (red points); theory (solid lines). The insets show the mask for $|\psi\rangle$ and the ``filtered'' masks for each $|\psi_\ell\rangle$. In the far-field plots, the red circles indicate the positions given by Eq.~(\ref{eq:positions}) where the ptychographic data are taken. Normalizations use the maximum intensities generated by (a) $|\psi\rangle$ and (b)--(g) $|\psi_\ell\rangle$ (in this case, $|\psi_2\rangle$).}
\end{figure*}

Our experimental setup is shown in Fig.~\ref{fig:setup}(c). We use a single-mode diode laser at $\lambda=687\!$~nm whose beam profile is spatially filtered, expanded and collimated. It is polarized in the vertical direction and normally incident on a reflective liquid crystal on silicon microdisplay (Holoeye PLUTO) working as a programmable phase-only spatial light modulator (SLM). The phase of this field is modulated by a computer-generated mask addressed at the SLM given by an array of blazed diffraction gratings. Typical masks are shown in the insets of Fig.~\ref{fig:NearFar}. For a display with pixels 8~$\!\mu$m wide, the gratings have period of 12 pixels, width and separation of 11 (9 and 5, respectively) pixels, for $d<20$ ($d\geq 20$). The modulated field is transmitted through the spherical lens $L_1$ and in its focal plane the first diffracted order is filtered by a slit diaphragm at both output arms of a beam splitter (BS). The filtered field emerges as a coherent superposition of $d$ spatial modes, where the magnitude and phase of each mode are controlled by the phase depth of the grating and its relative lateral displacement, respectively \cite{Prosser13,Varga14}. This field is isomorphic to a pure $d$-dimensional quantum state. The high level of purity of such states have been characterized in many other experiments, e.g.,  \cite{Goyeneche15,Stefano17,Stefano19,Prosser17,Prosser13}.  CMOS cameras (Thorlabs DCC1545M) at the transmitted and reflected arms record the far-field and near-field intensity distributions, respectively. 

The insets of Fig.~\ref{fig:setup}(c) show the role of each module in our setup. 
In the first module, we emulate the first step of ptychography ($\hat{P}_\ell|\psi\rangle$) by directly preparing the projected states $|\psi_\ell\rangle$ from the target state $|\psi\rangle$. Since the projectors from Eq.~(\ref{eq:Proj}) act as binary filters in $\mathcal{H}_d$ [each level either is or is not selected, as seen in Fig.~\ref{fig:setup}(b)], this corresponds to applying binary filters to the mask that generates $|\psi\rangle$.  In the other modules, the spatial intensity distributions associated with the incoming $|\psi_\ell\rangle$ and recorded by the cameras, are used to obtain the outcomes of the projective measurements in the computational (near-field) and Fourier basis (far-field).The lens $L_1$, shared between two modules, assists both preparation and measurement stages.

To illustrate the above discussion, Fig.~\ref{fig:NearFar} shows the (normalized) one-dimensional intensity distributions measured for a $d=6$ dimensional state. The near- and far-field measurements (red points) are plotted in the first and second rows, respectively; the solid curves are the theoretical predictions and the insets show the mask addressed to the SLM in each case. In Fig.~\ref{fig:NearFar}(a), we plot the results for the target state $|\psi\rangle$. Figures~\ref{fig:NearFar}(b) to \ref{fig:NearFar}(g) show the results for $|\psi_0\rangle$ to $|\psi_5\rangle$, respectively. Their near-field graphs show the correspondence between filtering the $|\psi\rangle$ mask or filtering $|\psi\rangle$ itself (in this case, with rank-3 projectors). The red circles in the far-field graphs indicate the transverse positions given by Eq.~(\ref{eq:positions}) where we take the ptychographic data.

We performed the experiment for several dimensions from $d=3$ to $d=32$.  For each $d$ we reconstructed a number of pure states randomly generated according to the Haar measure. Given a target state $|\psi\rangle$, we built its mask $M_{\psi}$, and from it,  the corresponding $n=d$ ``filtered'' masks $\{M_{\psi_\ell}\rightarrow|\psi_\ell\rangle\}_{\ell=0}^{d-1}$, following the prescriptions for the family (ii) of projectors [see discussion below Eq.~(\ref{eq:Proj}) and insets of Fig.~\ref{fig:NearFar}]. Each mask is addressed to the SLM one at a time and the camera images are recorded at near- and far-field. In both cases we take three images per mask, average over them, integrate over the transverse direction $y$, and subtract the background noise, obtaining one-dimensional patterns as seen in Fig.~\ref{fig:NearFar}. All far-field images are taken within the first single-slit diffraction minima. 

The postprocessed images for $M_\psi$ [e.g., Fig.~\ref{fig:NearFar}(a)] are used to characterize the preparation by taking the spatial mode magnitudes (the normalized intensities of the slits) as inputs into a least squares fitting of the entire far-field intensity distribution (see supplementary material of Ref.~\cite{Prosser17}). The retrieved source state, $|\psi_{\rm src}\rangle$, will be used later to evaluate the quality of the ptychographic reconstruction. 

For the ptychography, we use {\it only} the postprocessed far-field images for each $M_{\psi_\ell}$. In the $\ell$-th image we take {\it only} the intensities $I_{\ell j}$ at the $d$ pixels in the positions $x_j$ given by Eq.~(\ref{eq:positions}) (see also Fig.~\ref{fig:NearFar}). Gathering these intensities, the ptychographic data are settled as $\{\Pi_\ell=\{I_{\ell j}^{1/2}\}_{j=0}^{d-1}\}_{\ell=0}^{d-1}$. To perform the reconstruction with the family (ii) of projectors, we use all $\Pi_\ell$'s. To do the same with the family (i), we select the subset of five $\Pi_\ell$'s associated with the $M_{\psi_\ell}$'s which accomplish the prescriptions for that family [see below Eq.~(\ref{eq:Proj})]. In either case, the data is fed into the PIE algorithm [Fig.~\ref{fig:setup}(a) and description therein] and its normalized estimate, $|\phi_{\rm\textsc{pie}}\rangle$, is used to calculate the fidelity of the ptychographic reconstruction as $F=|\langle\phi_{\rm\textsc{pie}}|\psi_{\rm src}\rangle|^2$. 

\begin{figure*}[t!]
\centerline{\includegraphics[width=1\textwidth]{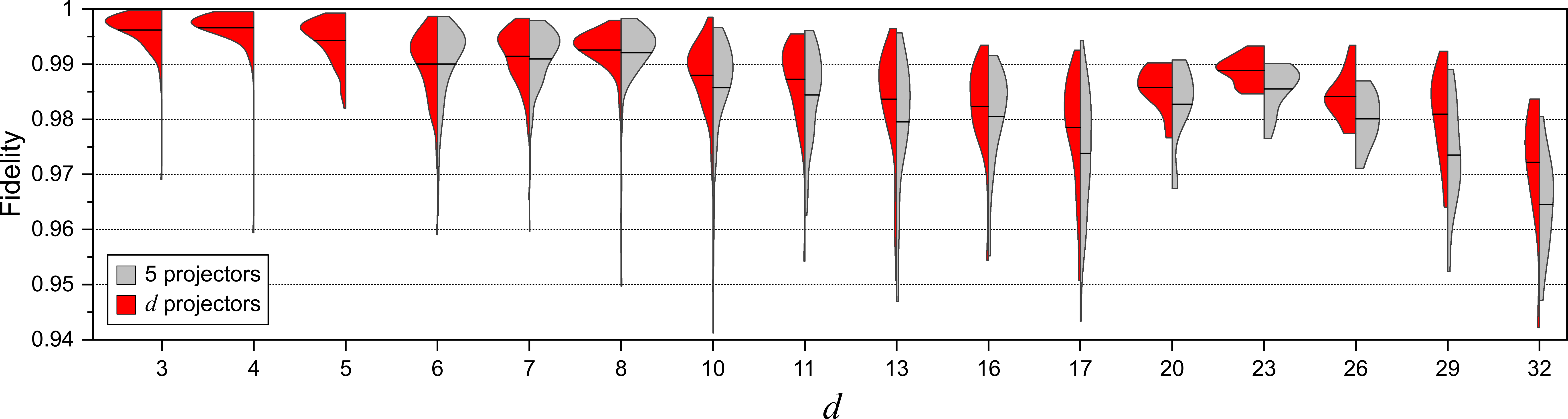}}
\caption{\label{fig:violin}  Split violin plots with the distribution of fidelities as a function of $d$ for the ptychographic reconstructions using $n=5$ and $n=d$ projectors given by Eq.~(\ref{eq:Proj}). The horizontal solid lines represent the average fidelities. See text for more details. }
\end{figure*}

For $d\leq 10$, $11\leq d\leq 17$, and $d>17$, we reconstructed 100, 50, and 13 random states per dimension, respectively. The obtained fidelities are shown in the split violin plots of Fig.~\ref{fig:violin}, where the distributions in gray (red) correspond to the ptychography with $n=5$ ($n=d$) projectors. For a better visualization, we scaled the width of each violin by  area, i.e., each one will have the same area regardless of the number of reconstructions. The horizontal solid line cutting each violin represents the average fidelity. As $F\in[0,1]$ and $F=1$ characterizes a perfect reconstruction, the fidelities achieved for all dimensions and both families of projectors were consistently high, showing that all states prepared by the source were faithfully reconstructed by the pytchographic method.  

Considering only the state-space dimension, Fig.~\ref{fig:violin} shows that, in general, the fidelities decrease as $d$ increases. This trend is related with the growing experimental imperfections in the preparation and measurement stages. An increasing $d$ decreases the purity of the source states since the coherence between the transverse spatial modes is affected. Note that for $d=20$ the trend is broken and restarted at a higher level because, as described earlier, we reduced the width and separation of the spatial modes (for $d\geq 20$), thus reducing decoherence effects. At the measurement stage, the errors in the detection process grow with the number of Fourier basis states, $d$.

The above discussion concerns our particular implementation and not the ptychographic method in general. Let us now analyze it in more details by comparing the reconstructions with the two families of projectors $\{\hat{P}_\ell\}_{\ell=0}^{n-1}$ used here. As seen in Fig.~\ref{fig:violin} for $d\geq 6$, the dispersion of both fidelity distributions for each $d$ are similar while the averages for $n=d$ are greater than for $n=5$, which becomes more evident as $d$ increases. These results show that, on average, using more projectors leads to better reconstructions since the ptychographic data will have more diversity and redundancy, the key elements for the method. On the other hand, using less projectors implies in less measurement settings and, as we observed here, provides reconstructions with comparable quality which can be further improved by employing more robust versions of the PIE algorithm \cite{Maiden17}. 

For $n=5$, the postprocessing times, $t_{\rm\textsc{pie}}$, took few milliseconds per state $\forall d$; for $n=d$, $t_{\rm\textsc{pie}}\approx d^3$, but did not last more than 230 milliseconds for $d=32$ on a modest laptop (see Ref.~\cite{Fernandes19}).

The present form of quantum ptychography is resilient to noise and works quite well for state purities up to $\sim 90$\% \cite{Fernandes19}. Below that, in general, the reconstruction algorithm fails to converge. Thus, our results confirm the high level of purity of the source states. The generalization of the method to mixed states will require different specifications for the ``slicing'' operators [Eq.~(\ref{eq:Proj})] and also in the update rule of the algorithm.

We can straightforwardly extend our implementation to other types of  photonic spatial modes. For instance, if we consider pure states encoded into angular or longitudinal spatial modes, the rank-$r$ projectors of Eq.~(\ref{eq:Proj}) would be implemented with simple mode blockers, as discussed earlier. For the angular states, the measurement in the Fourier basis would be performed through an optical Fourier transform \cite{Yao06}, whereas for the longitudinal states it would be carried out with a multiport integrated photonic device \cite{Crespi16}.

We performed a proof of concept demonstration of quantum state ptychography. The method faithfully reconstructed pure states in dimensions up to 32, and it was shown to be simple and flexible in regard to the measurement settings, and robust against noise. Like ptychography, quantum ptychography may be generalized in many ways; here we just demonstrated its first application.

\medskip
\noindent\textbf{Funding.} FAPEMIG (APQ-00240-15); CNPq INCT-IQ (465469/2014-0); CNPq (407624/2018-0, 140359/2017-6); Universidad de La Frontera (DI20-0154).

\medskip
\noindent\textbf{Disclosures.} The authors declare no conflicts of interest.

\bibliography{REFS}

\end{document}